\documentclass[showpacs,preprint,amsmath,amssymb]{revtex4}
\usepackage{graphicx}% Include figure files
\usepackage{dcolumn}% Align table columns on decimal point
\usepackage{bm}% bold math
\begin{document}

%\preprint{APS/123-QED}
%\begin{center}

\title{Magnetic models on Apollonian networks}
\author{Roberto F. S. Andrade}
\affiliation{
Instituto de F\'{i}sica,\\
Universidade Federal da Bahia,\\
}
\author{Hans J. Herrmann}
 \altaffiliation[Also at ] {Institute for Computerphysics,\\
 University of Stuttgart, Germany\\
email: hans@ica1.uni-stuttgart.de }
 \affiliation{Departmento de F\'{i}sica,\\
Universidade Federal do Cear\'{a},\\
 60450 Fortaleza, CE, Brazil}

\date{\today}% It is always \today, today,

%\vspace{0.2in}
%\end{center}

\begin{abstract}
Thermodynamic and magnetic properties of Ising models defined on
the triangular Apollonian network are investigated. This and other
similar  networks are inspired by the problem of covering an
Euclidian domain with circles of maximal radii. Maps for the
thermodynamic functions in two subsequent generations of the
construction of the network are obtained by formulating the
problem in terms of transfer matrices. Numerical iteration of this
set of maps leads to exact values for the thermodynamic properties
of the model. Different choices for the coupling constants between
only nearest neighbors along the lattice are taken into account.
For both ferromagnetic and anti-ferromagnetic constants, long
range magnetic ordering is obtained. With exception of a size
dependent effective critical behavior of the correlation length,
no evidence of asymptotic criticality was detected.

\end{abstract}

\maketitle

\section{Introduction}

The investigation of magnetic models on scale invariant networks
has attracted the attention of scientists since the 80's
\cite{Gefen80, Gefen84}. Besides the fact that, on such graphs,
renormalization procedures can lead to exact results\cite{MK},
they have been explored as models for systems that are not
translational invariant, neither in the positions of the spins nor
in the coupling constants mediating the interactions between them.
In this respect, the analysis of disordered and aperiodic models
on scale invariant graphs, which include hierarchical
lattices\cite{Berker,Kauffman,Tsallis}, Cayley trees \cite{Yokoi}
or Sierpinski gaskets and carpets\cite{Luscombe,Bonnier}, have
provided valuable insight into the behavior of critical phenomena
of non homogeneous systems on Euclidean lattices.

A further family of scale invariant graphs are Apollonian
networks, the simplest of which is illustrated in Fig. 1. This
lattice can be defined based of the ancient problem of filling
space with spheres, first tackled by the Greek mathematician
Apollonius of Perga \cite{Boyd73}. In its two dimensional version,
corresponding to the problem of the plane filled by circles, the
nodes of this network are defined by the positions of the centers
of the circles, while edges are drawn between any pair of nodes
corresponding to pairs of touching circles\cite{Herrmann90}. The
resulting network, corresponds to the contact force network of the
packing\cite{Mahmoodi04}. Apollonian networks can also be used to
describe generically other scale-free situations like
space-filling porous media \cite{Dodds80} or the connections
between densely located cities on which one is interested in fluid
flow, car traffic or electric supply. Therefore it is useful to
study not only the geometric properties of Apollonian networks but
also transport and ordering on them.

In a previous paper \cite{Andrade04}, we analyzed several of its
properties. In particular, we have shown that it has small-world
properties, a scale-free degree distribution, a very high
clustering coefficient and a very short diameter, all this having
been confirmed independently by Doye \cite{Doye04}. Moreover, the
Apollonian network can be embedded in the Euclidian plane, what is
not the case for other scale invariant lattices, e.g. hierarchical
lattices or Cayley trees. One can of course define other similar
lattices based on modified packing rules \cite{Oron00}. Although
the numerical values for exponents depend on the topology of each
realization, basic properties characterizing complex networks
remain the same.

In our first work, we have also devoted our attention to several
physical models on the Apollonian network (electrical resistance,
percolation, magnetic ordering), pointing out the most striking
features. In this work, we review our investigations on the
properties of several Ising models on the Apollonian network, and
present a thorough discussion, regarding on one hand some of the
details of the Transfer Matrix (TM) methods and, on the other
hand, the most important thermodynamic and magnetic properties.
Several choices for rules defining the values of the coupling
constants are considered, including both ferro (F)- and
anti-ferromagnetic (AF) interactions. As will be shown, we found
long range magnetic ordering for almost all choices of couplings
without any noticeable evidence of a phase transition to a
paramagnetic phase at a finite temperature. For particular choices
of F and AF bonds, the geometry of the network induces the
presence of competition and frustration within closed loops of odd
number of sites, giving rise to residual entropy and changes in
the correlation length. The behavior of this quantity deserves a
detailed discussion, as it points to a transition from long to
short range correlation only for finite size systems, in a similar
way as discussed for magnetic models on scale-free
lattices\cite{Stauffer02}

The rest of this paper is organized as follows: in Section II we
discuss our models. In Section III we obtain the maps for the free
energy and its derivatives, as well as for the correlation length.
Results are discussed in Section IV, while concluding remarks are
presented in Section V.

\section{Ising models}

The Apollonian network is constructed recursively. In each
generation, it incorporates a new set of sites, which correspond
to the centers of the new circles added to the packing filling the
holes left in the previous generation. In the present work we
consider the lattice which starts with three touching circles
drawn on the vertices of an equilateral triangle, and the packing
problem is restricted to filling the space bounded by these three
initial circles, as shown in Figure 1a. If $n$ denotes the current
generation of the network, the number of sites $N(n)$ is
asymptotically three times that of the previous generation, i.e.
$N(n+1)=3N(n)-5$, or $N(n)=(3^{n-1}+5)/2$. The number of edges
linking nodes increases with $n$ according to
$B(n+1)=B(n)+3(N(n+1)-N(n))$. As a consequence,
$B(n)=(3+3^{n})/2$, $B(n)/N(n)\rightarrow3$ in the limit of large
$n$, so that on average, each site is linked to six other sites,
which is the coordination number of the triangular lattice.

\begin{figure}
\begin{center}
\includegraphics*[width=6.5cm,angle=270]{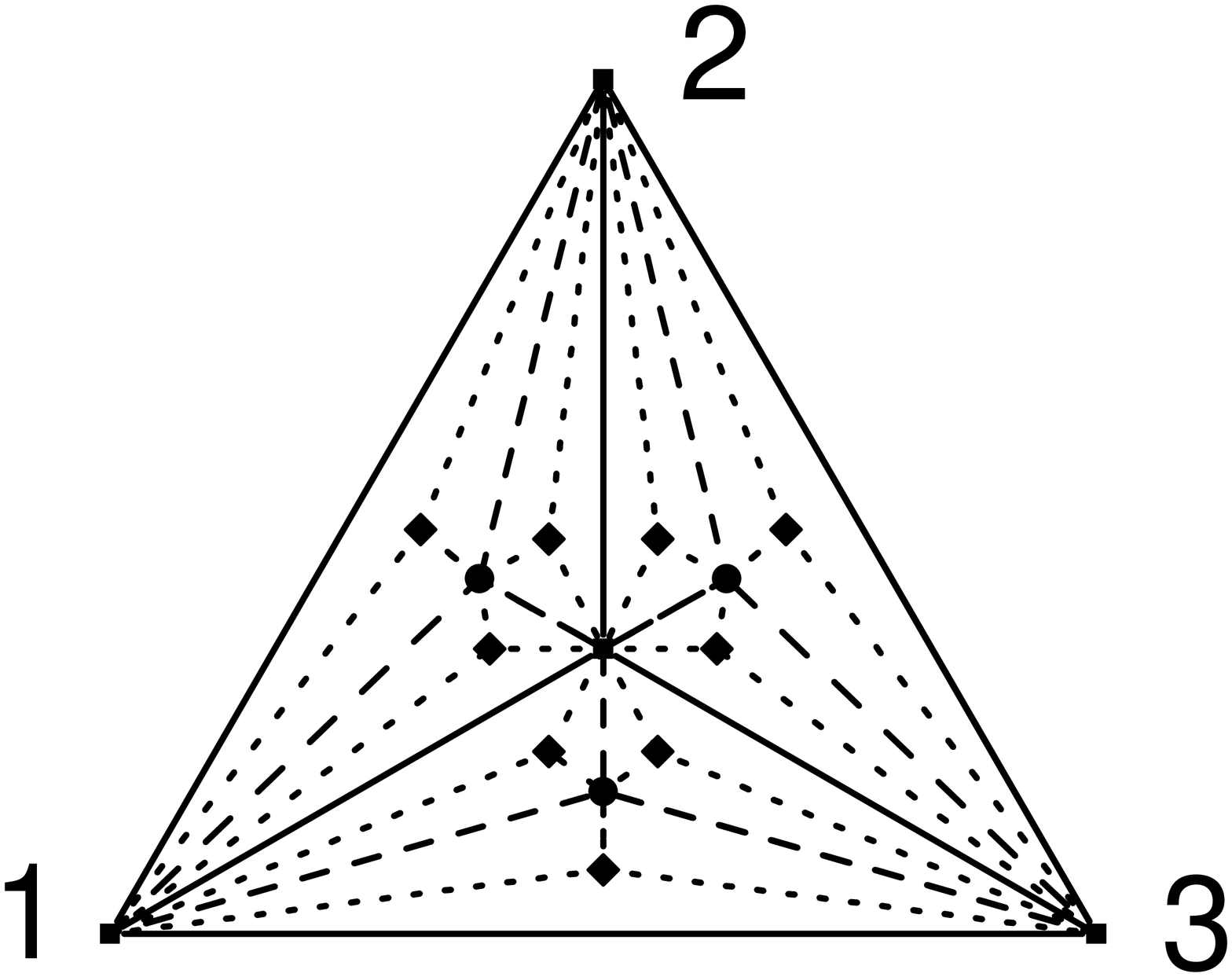}
\end{center}
\caption{Third generation of construction of the Apollonian
network. Sites represented by squares, circles and diamonds are
introduced in the first, second and third generation. Links
represented by dotted, dashed and solid lines correspond to $n=1,2$
and $3$ in eq(\ref{eq8}).} \label{fig1}
\end{figure}

Once the Apollonian network has been defined, it is possible to
define many different models on it. In this work we focus on a set
of interacting Ising spins $s_i=\pm1$  placed on each site of the
network. Interactions are restricted to pairs of spins placed on
nodes linked by edges, as described in the Introduction. So, some
spins placed far apart may interact, while pairs of spins placed
relatively close to each might not. This can be of interest to
model special disordered systems, having interactions of all
ranges.

For the packing problem, it is important to calculate the radius
of each circle $c(i,n)$, which depends both on the generation $n$
as well as on the local environment, i.e., the generations of the
three circles that it touches $n_p(i,k),k=1,2,3$. It is possible
to include this dependence into the magnetic model, by defining
coupling constants (or local fields) depending on the distance
between nodes, i.e., $J(n,i,n_p(i,j))$.  We restrict ourselves,
however, to a simpler situation, where the coupling constants $J$
only depend on the generation $n$ at which the edge was introduced
into the network. In Figure 1b we illustrate the first three steps
of the construction of the model.

To have a physically interesting model, it may be reasonable to
choose values for $J$ that increase with $n$. Indeed, when $n$
increases, the average length of the edges introduced in that
generation decreases and, as the spins get closer, we might expect
that the interactions among them become stronger. However, to
avoid the divergence of the energy within the lattice, it is
necessary to renormalize the value of all $J$'s as $n$ increases.
To accomplish this, we define $J_{n,m}, m=1,...,n$ as the value of
the constant introduced in the $m-th$ generation, when the lattice
has been built up to its $n$-th generation, and require that
$J_{n,m}$ decreases with $n-m$. A sufficiently general choice
would be
\begin{equation}\label{eq1}
J_{n,m}=\frac {(\pm1)^{m}J_{0}} {(n-m+1)^{\alpha}},
\end{equation}
where $J_0$ may have ferro- ($>0$) or antiferromagnetic ($<0$)
character, the exponent $\alpha$ controls how the interactions
decay with the difference $n-m$, and yields a possibility of
choosing the interactions according to the generation at which
they were introduced.

One of the extreme situations, $\alpha=0$, corresponds to equal
interactions along all edges in the network. On the other hand, in
the $\alpha\rightarrow\infty$ limit, the model contains only
finite interactions for the subset of edges that were introduced
in that last generation $n$, as illustrated in Figure 1b. The
number of these surviving bonds is given by $B(n)-B(n-1)=3^{n-1}$,
so that the average coordination number is reduced to four.
Moreover, the lattice is then composed by four-sided polygons, so
that competition and frustration due to the presence of
antiferromagnetic bonds can never occur in this $\alpha$ limit.

To close this section we write down the formal Hamiltonian of this
model
\begin{equation}\label{eq2}
H_n = -\sum_{(i,j)}\overline{J_{i,j}}s_i s_j - h \sum_{i} s_i,
\end{equation}
where all pairs of nearest neighbors denoted by $(i,j)$ are
defined according to the construction rules of the network, and
the  constants $\overline{J_{i,j}}$ must be chosen from the set
defined in eq.(\ref{eq1}) according to the value of $m$ in which
the edge was introduced. We also include a constant magnetic field
$h$, which allows for the evaluation of magnetic properties. The
notation in eq.(\ref{eq2}) does not include the selection of the
bonds that are taken into account. The evaluation of a partition
function can in fact be set up in much proper terms through the
transfer matrix formalism.

\section{Transfer matrix and recurrence relations}

The numerical evaluation of the partition function for magnetic
models on scale invariant graphs with a finite number of end nodes
has been performed with the help of TM derived maps for a large
number of lattices and models. The problem we are interested in
this work is also suitable to be analyzed within this framework.
For the sake of simplicity, let us first consider the homogeneous
case $\alpha=0$, and set $h=0$. If we consider the first
generation $n=1$, we observe that a $2\times2$ TM $M_1$, which
takes into account all interactions between the sites $i$ and $k$
of Figure 1 can be written as:
\begin{equation}\label{eq3}M_1=
\begin{pmatrix}
 a_{1} & b_{1} \\
 b_{1} & a_{1}
\end{pmatrix}=
\begin{pmatrix}
  a(a^{2}+b^{2}) & 2ab^{2} \\
  2ab^{2} & a(a^{2}+b^{2})
\end{pmatrix},
\end{equation}
where $a=b^{-1}=\exp(\beta J_0)$. $M_1$ can be used to describe
one single cell or a linear chain of triangles that are connected
by  their bases. It is also possible to define a $2\times4$ TM
$L_1$, that describes the interactions among sites $i,j$ and $k$,
where the column labels $\kappa$ are composed from the pair
$(j,k)$ according to the lexicographic order, i.e.,
$\kappa=2(j-1)+k$ according to
\begin{equation}\label{eq4}L_1=
\begin{pmatrix}
 c_{1} & d_{1} & d_{1} & d_{1} \\
 d_{1} & d_{1} & d_{1} & c_{1}
\end{pmatrix}=
\begin{pmatrix}
  a^{3} & ab^{2} & ab^{2} & ab^{2} \\
  ab^{2} & ab^{2} & ab^{2} & a^{3}
\end{pmatrix}.
\end{equation}
Of course we note that  $a_{1}=c_{1}+d_{1}$ and $b_{1}=2d_{1}$.

Within the proposed framework, all interactions between the sites
$i$ and $k$, for any higher order generations $n=2,3,4...$, should
be written in terms of a single $2\times2$ TM's $M_n$, with the
same distribution of matrix elements as $M_1$. Moreover, the
matrix elements of $M_n$ should be written in terms of those of
the matrices of the lower generation $n-1$ only. This turns out to
be feasible since the Apollonian lattice, in a generation $n+1$,
can be decomposed into three sublattices, each one of them being a
deformed lattice of generation $n$. Since the coupling constants
do not depend on the actual distance between the sites, each of
the three sublattices entails the same coupling constants and
magnetic structure as the $n$-lattice. Thus, a matrix $M_{n+1}$
can indeed be written in terms of three matrices $M_n$. To achieve
this we remark that, in any generation, the three sublattices
share their three outmost sites, which we label as $i,j,k$ and
$\ell$. This last one
 occupies the geometrical center of the
$n+1$-lattice. Then, $M_{n+1}$ can be defined using
\begin{equation}\label{eq5}
(M_{n+1})_{i,k} = \sum_{j,\ell}(L_n)_{i,j\ell} (L_n)_{i,\ell k}
(L_n^t)_{k,j\ell}.
\end{equation}
and
\begin{equation}\label{eq6}
(L_{n+1})_{i,jk} = \sum_{\ell}(L_n)_{i,j\ell} (L_n)_{i,\ell k}
(L_n^t)_{k,j\ell}.
\end{equation}

As one can easily observe by direct evaluation of eq.(\ref{eq5})
and eq.(\ref{eq6}), all matrices $M_n$ and $L_n$, share the same
matrix element distribution as $M_1$ and $L_1$. So, it is possible
to immediately write down recurrence relations for the elements of
$L_{n+1}$ in terms of those of $L_{n}$ as
\begin{equation}\label{eq7}
\begin{array}{c}
 c_{n+1} = c_n^3 + d_n^3\\
 d_{n+1} = c_nd_n^2 + d_{n}^3
\end{array},
\end{equation}
from which the elements $a_{n+1}=c_{n+1}+d_{n+1}$ and
$b_{n+1}=2d_{n+1}$ can be obtained.

However, a direct evaluation of the matrix elements defined by the
equations (\ref{eq5})-(\ref{eq7}) shows that they do not exactly
describe the interactions between the sites $i$ and $k$. For
instance, in the generation $n=2$, the number of magnetic bonds is
equal to the number of edges $B(n=2)=6$. On the other hand, we see
that the Boltzmann weights in $M_2$ are expressed by combinations
of $\exp(\beta rJ_0)\exp(-\beta sJ_0)$, with $r+s=9$ instead of
$r+s=6$. This is due to the fact that each one of the interactions
between the site $\ell$ and its neighbors $i,j$ and $k$ appears
twice in eq.(\ref{eq5}). To describe the thermodynamics of the
system with the help of equations (\ref{eq5})-(\ref{eq7}), it is
necessary to carry out a small correction, namely to redefine $a$
and $b$ as $a=b^{-1}=\exp(\beta J_0/2)$. With this modification,
the Boltzmann weights in each element are expressed by
exponentials of $\beta (r-s)J_0$, where $r+s=B(n)-3/2$. So, for
each spin configuration, the ratio between the correct energy and
that one provided by equations (\ref{eq5}) and (\ref{eq6}) is
roughly proportional to $(B(n)-3/2)/B(n)$, which $\rightarrow1$ in
the limit $n\rightarrow\infty$.

These definitions are sufficient for  only  the uniform
interaction model $\alpha=0$. Further modifications in
eqs.(\ref{eq5})-(\ref{eq7}) are required to obtain the correct
maps for general $\alpha$. To cast this into a single recurrence
relation, we first note that it is not necessary to use two labels
$n$ and $m$ to insert the correct coupling constants into $M_n$.
As these matrices are recursively defined, the largest and most
abundant $J_{n,n}$ corresponds to the constant introduced into
$M_1$, which is reproduced in ever growing number by the
successive use of equations like (\ref{eq5})-(\ref{eq6}). On the
other hand, the smallest and least frequent $J_{n,1}$ represents
the constant that is inserted into the sequence of TM's exactly at
the $n$-th generation. So we consider
\begin{equation}\label{eq8}
J_{n,m}\rightarrow\mathbf{J_{n}}=\frac {(\pm1)^{n}J_{0}}
{(n-1)^{\alpha}}.
\end{equation}
Note that the changes carried into the denominator of
eq.(\ref{eq8}) set $\mathbf{J_{1}}=0$, and
$\mathbf{J_{n+1}}\rightarrow\mathbf{J_{n}}$. This strategy is
necessary to avoid taking into account more than once the effect
of bonds introduced when $n=1$, and it is somehow equivalent to
the redefinition of $a$ and $b$ discussed above.

Then we modify equations (\ref{eq5}) and (\ref{eq6}) according to
\begin{equation}\label{eq9}
(M_{n+1})_{i,k} = \sum_{j,\ell}(L_n)_{i,j\ell} (L_n)_{i,\ell k}
(L_n^t)_{k,j\ell} (C_n)_{i,\ell} (C_n)_{\ell,j} (C_n)_{\ell,k}
\end{equation}
and
\begin{equation}\label{eq10}
(L_{n+1})_{i,k} = \sum_{\ell}(L_n)_{i,j\ell} (L_n)_{i,\ell k}
(L_n^t)_{k,j\ell} (C_n)_{i,\ell} (C_n)_{\ell,j} (C_n)_{\ell,k},
\end{equation}
where the $2\times2$ TM´s $C_n$ are defined by
\begin{equation}\label{eq11}C_n=
\begin{pmatrix}
 p_{n} & q_{n} \\
 q_{n} & p_{n}
\end{pmatrix}=
\begin{pmatrix}
  \exp(\beta \mathbf{J_n}) & \exp(-\beta \mathbf{J_n}) \\
  \exp(-\beta \mathbf{J_n}) & \exp(\beta \mathbf{J_n})
\end{pmatrix}.
\end{equation}
With these definitions, it is possible to observe that the number
of non-zero coupling constants in the lattice is $B(n)-3$ so that,
in the $n\rightarrow\infty$ limit, equations
(\ref{eq8})-(\ref{eq11}) accurately describe the thermodynamic
properties of the model. The recurrence maps for the matrix
elements derived from (\ref{eq9})-(\ref{eq10}) read
\begin{equation}\label{eq12}
\begin{array}{c}
 c_{n+1} = c_n^3p_n^3 + d_n^3q_n^3\\
 d_{n+1} = c_nd_n^2p_n^2q_n + d_{n}^3p_nq_n^2
\end{array},
\end{equation}

From equations (7) or (11) it is possible to derive recurrence
maps for the free energy $f_n=-T\ln(c_n+d_n)/N(n)$ and correlation
length $\xi_n=1/\ln((c_n+d_n)/(c_n-d_n))$ at two subsequent
generations, $f_{n+1}=f_{n+1}(f_n,\xi_n;T)$ and
$\xi_{n+1}=\xi_{n+1}(f_n,\xi_n;T)$ can easily be derived
\cite{Kohmoto83,Andrade99}. This set of maps can be increased by
working out explicit recurrence relations for the derivatives of
$f_n(T)$ with respect to both the temperature and the magnetic
field, obtaining the entropy $s(T)$, the specific heat $c(T)$, the
spontaneous magnetization $m(T)=m(T,h=0)$, and the magnetic
susceptibility  $\chi(T,h=0)$. For this last purpose, we have to
consider $h\neq0$ and insert it into the matrices $M_n$. This
modification breaks the up-down symmetry of the problem, so that
the matrices $M_n$ and $L_n$ have a larger number of distinct
matrix elements. This is a straightforward procedure that has been
carried out for other models \cite{Kohmoto83,Andrade99}. In the
Appendix we present the full set of recurrence maps used in this
work.

\section{Results}

We study the thermodynamic functions, i.e. the free energy $f$,
the entropy $s$, the specific heat $c$, the spontaneous
magnetization $m$ and the correlation length $\xi$ as function of
the temperature $T$, as shown in Figures 2 to 5. They were
obtained by numerically iterating the set of maps shown in the
Appendix, starting with $T$ dependent initial conditions, until a
value $n^{*}(T)$. This temperature dependent value is set
automatically, by requiring that one (or a set) of the intensive
quantities and/or the correlation length, have converged to a
fixed value, within a previously established relative tolerance.
This is usually $\sim10^{-15}$, as we work with double precision
variables. Convergence based only on the value of $f$ and its
derivatives is much faster than for $\xi$, specially when the
system is in the ordered phase. Otherwise stated, if we call
$n_{x}^{*}(T),x=f,s,c,m,\xi$, the value of $n$ at which the
function $x$ has converged for that particular value of $T$, then
we find that $n_{\xi}^{*}(T)$ always assumes  the largest value.

\begin{figure}
\begin{center}
\includegraphics*[width=11cm,angle=270]{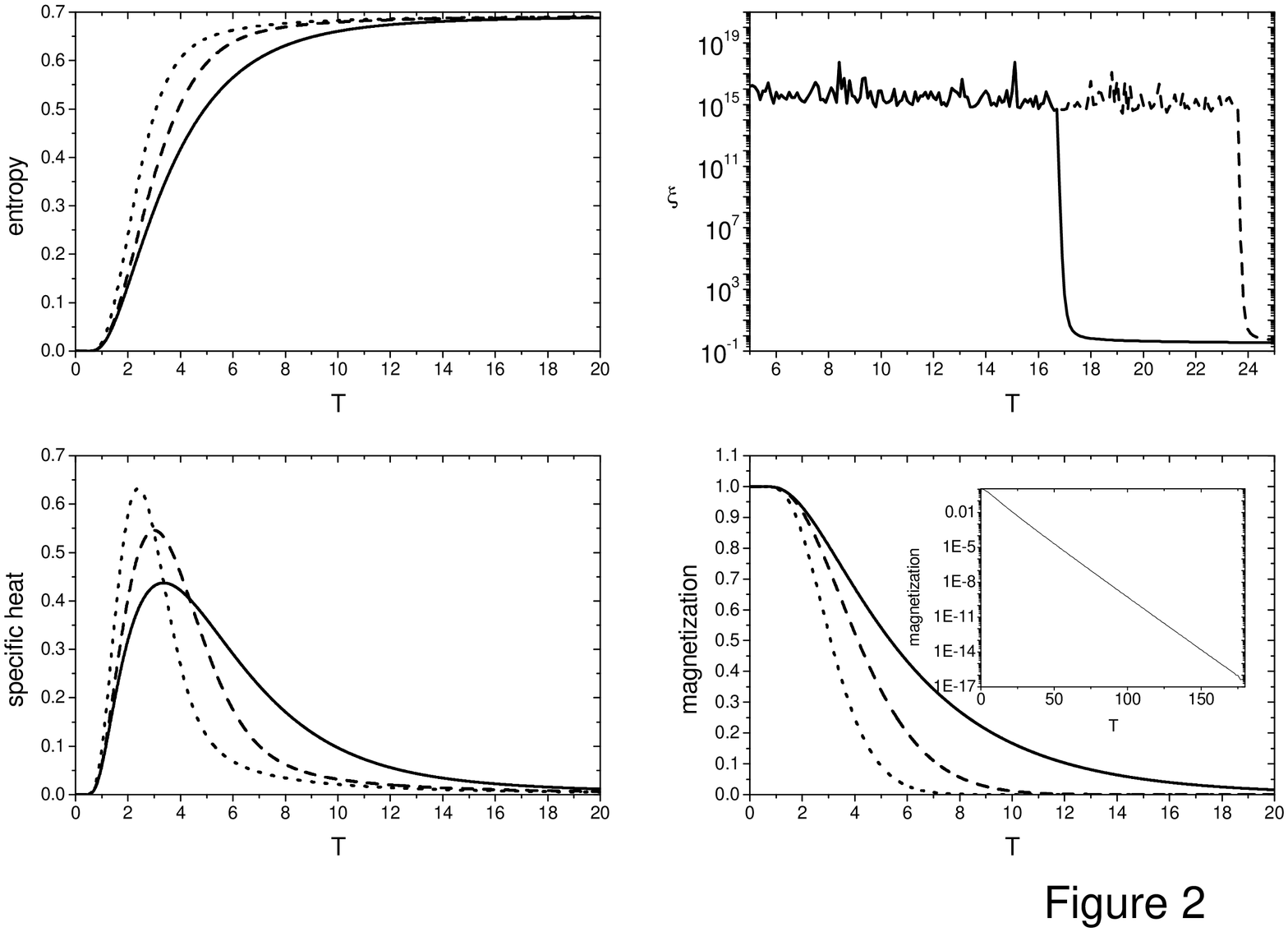}
\end{center}
\caption{Thermodynamic functions for the ferromagnetic model.
Solid, dashed and dotted lines indicate $\alpha=0,1$ and $\infty$.}
\label{fig2}
\end{figure}

In Figure 2 $a, b, c$ and $d$ we show the entropy $s$, the
specific heat $c$, spontaneous magnetization $m(h=0)$ and
correlation length $\xi$, for three distinct values of $\alpha,$
when all coupling constants have ferromagnetic character, i.e.,
$J_0>0$ and $(+1)^n$ in eq.(\ref{eq8}). The qualitative behavior
does not depend on the values of $\alpha (0, 1,$ and $\infty)$,
i.e., whether interactions are only short
$(\alpha\rightarrow\infty)$ or long range $(\alpha=0)$. For all
cases we see that, for low values of $T$, long range correlation
sets in, as is evident from the spontaneous magnetization and the
numerical divergence of $\xi$. The remarkable feature, however,
points to the absence of any criticality when $T$ is increased.
The insert in Fig 2b shows that, when $\alpha=0$, $m$ goes to zero
smoothly, as $\exp(-T)$, with no evidence of a sharp transition to
$m=0$ at a well defined critical temperature. If we consider
$\alpha>0$, we still find a smooth, but stronger, decay, namely as
$m\sim\exp(-T^\lambda)$. The curve for the specific heat is also
smooth, showing a typical Schottky maximum, again without any
evidence of a divergence, that would be expected for a usual phase
transition.

\begin{figure}
\begin{center}
\includegraphics*[width=6.5cm]{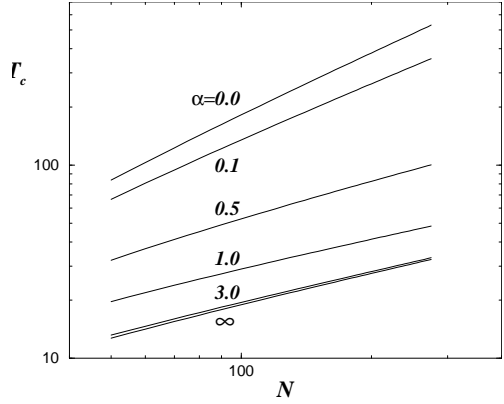}
\end{center}
\caption{Size dependent critical temperature for distinct values
of $\alpha$.} \label{fig3}
\end{figure}

The results for the correlation length (d) are also distinct from
those found for other scale invariant models, as the diamond
hierarchical lattice (DHL)\cite{Andrade99}. There, $\xi$ is finite
for large values $T$ and numerically diverges for all values of
$T$ below a well defined critical value $T_c$ which in our case
means attaining a value larger than $10^{16}$, the largest allowed
number in our algorithm. Within this region, the actual value
reached by $\xi(T)$, has no precise meaning. Typically it is much
higher than those in the disordered phase, and is also
characterized by the presence of random fluctuations. As mentioned
before, $n_{\xi}^{*}(T)$ is larger than $n_{x}^{*}(T),x=f,s,c,m$,
but even if we stop the iterations at $n_{f}^{*}(T)$, $\xi$ has
already reached this very high plateau. This shows that it is not
actually necessary to proceed further with the iteration of the
maps, as we would obtain only a meaningless value for $\xi$.

In the present case, if we use $n_{f}^{*}(T)$ to stop the
iteration of the maps, we observe that  $\xi$ diverges at low
temperatures, expressing long range order. When $T$ is increased
beyond a given value of $T^{*}$, it converges to a well defined
value, suggesting the break of long range correlation. However, if
the iteration procedure is pursued to a value of $n>n_{f}^{*}(T)$
we observe that the  $T$ interval in which  $\xi$ diverges becomes
larger. This finding has driven us to proceed with the iteration
of the maps in a different way. We fix a value
$\overline{n}>n_{f}^{*}(T=1)$, and iterate
 the maps until reaching $\overline{n}$ for all values within a large
$T$ interval, as shown in Figure 2c. Then it is possible to
precisely evaluate a critical value $T_c(\overline{n})$, as the
value of $T$ where the behavior of $\xi$  changes. In  Figure 3 we
show how $T_c(\overline{n})$ depends on $\overline{n}$, for
several distinct values of $\alpha$. Our findings for this unusual
kind of critical behavior suggest a power law $T_c(n)\sim
n^{\tau(\alpha)}$, with $\tau$ going continuously from
$\tau(\alpha=0) = 1$ to $\tau(\alpha=\infty) = 1/2$ . We recall
that a similar behavior has been reported for spin models on
another scale-free lattice \cite{Stauffer02}.

\begin{figure}
\begin{center}
\includegraphics*[width=11cm,angle=270]{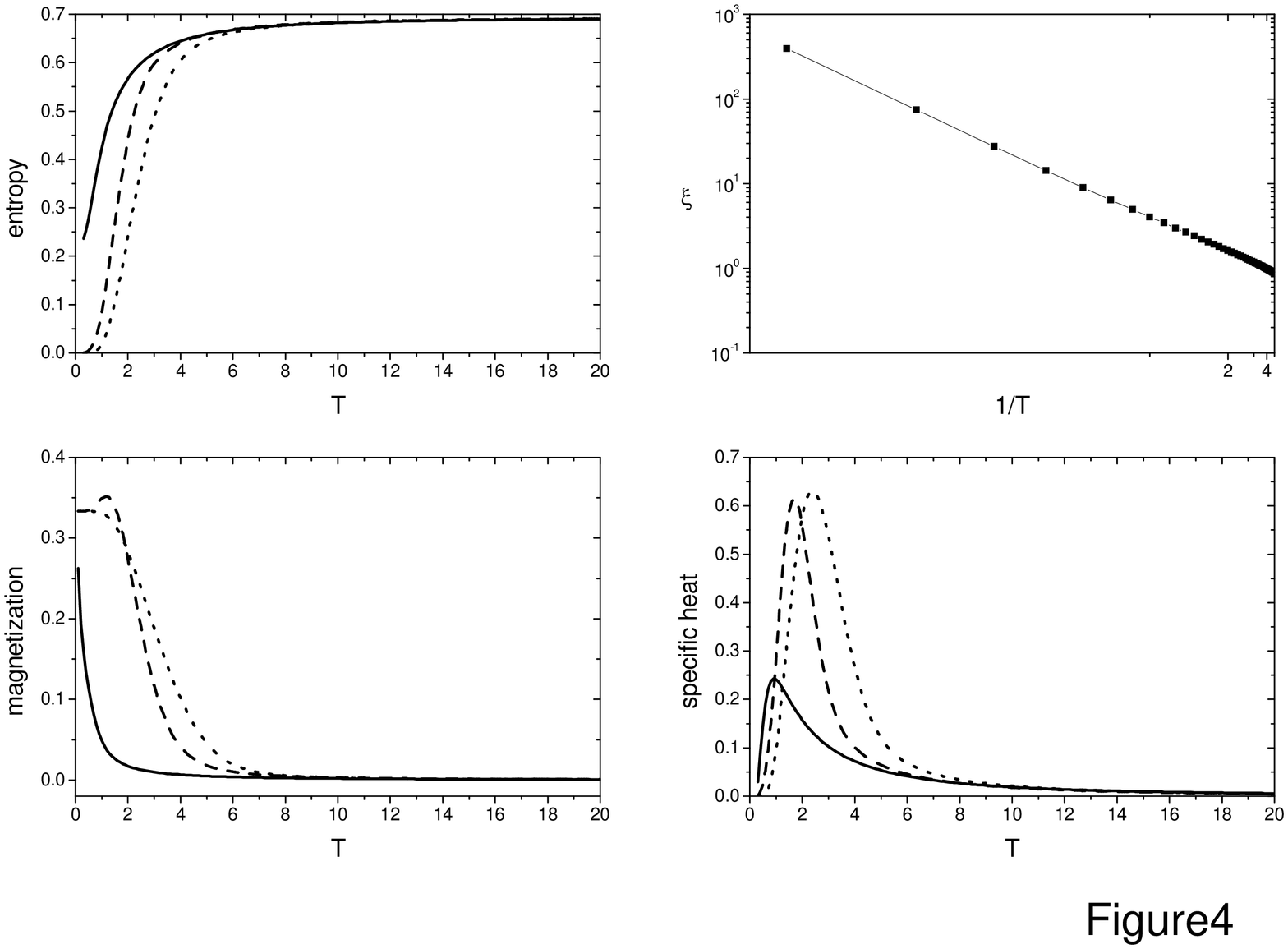}
\end{center}
\caption{Thermodynamic functions for the anti-ferromagnetic model.
Solid, dash and dots indicate $\alpha=0,1$ and $\infty$. $\xi$ is
drawn only for $\alpha=0$ and reciprocal temperature axis.}
\label{fig4}
\end{figure}

In Figure 4 we show that antiferromagnetic interactions $(J_0<0)$
change the thermodynamic behavior of the model. The most
interesting situation is observed for $\alpha=0$. All triangles in
the lattice are frustrated and, as expected, a residual entropy
$s_0=0.222$ is measured. This best numerical value is smaller than
that for the triangular lattice $s_0=0.3238..,$ \cite{Wannier},
and much smaller than that obtained for the Ising model on the
Sierpiski gasket $s_0=0.493..,$
(SG)\cite{Moreira,Stinchcombe,Andrade93}.

At the same time we find finite well defined values for $\xi$ for
all values of $T$, which are robust with respect to the value of
$\overline{n}$ where the iterations are stopped. This is
illustrated in Figure 4d, which also shows that, as
$T\rightarrow0$, $\xi$ decays like $\exp(-1/T)$, typical for the
1-d chain. This is a somewhat unexpected behavior, as the presence
of frustration usually does not allow for long range correlation
of spin orientation, even at $T=0$ (e.g. the AF Ising model on the
SG \cite{Andrade93}). At the same time, this result must be
related to the non-vanishing behavior for the magnetization, shown
in Figure 4c. It indicates that the number of spins pointing in
each direction is not the same. Once again, this behavior is
different from that obtained for other frustrated lattices, like
the planar triangular lattice or the SG. Finally, the behavior of
the specific heat looks like those found when $J_0>0$, for any
value of $\alpha.$

\begin{figure}
\begin{center}
\includegraphics*[width=11cm,angle=270]{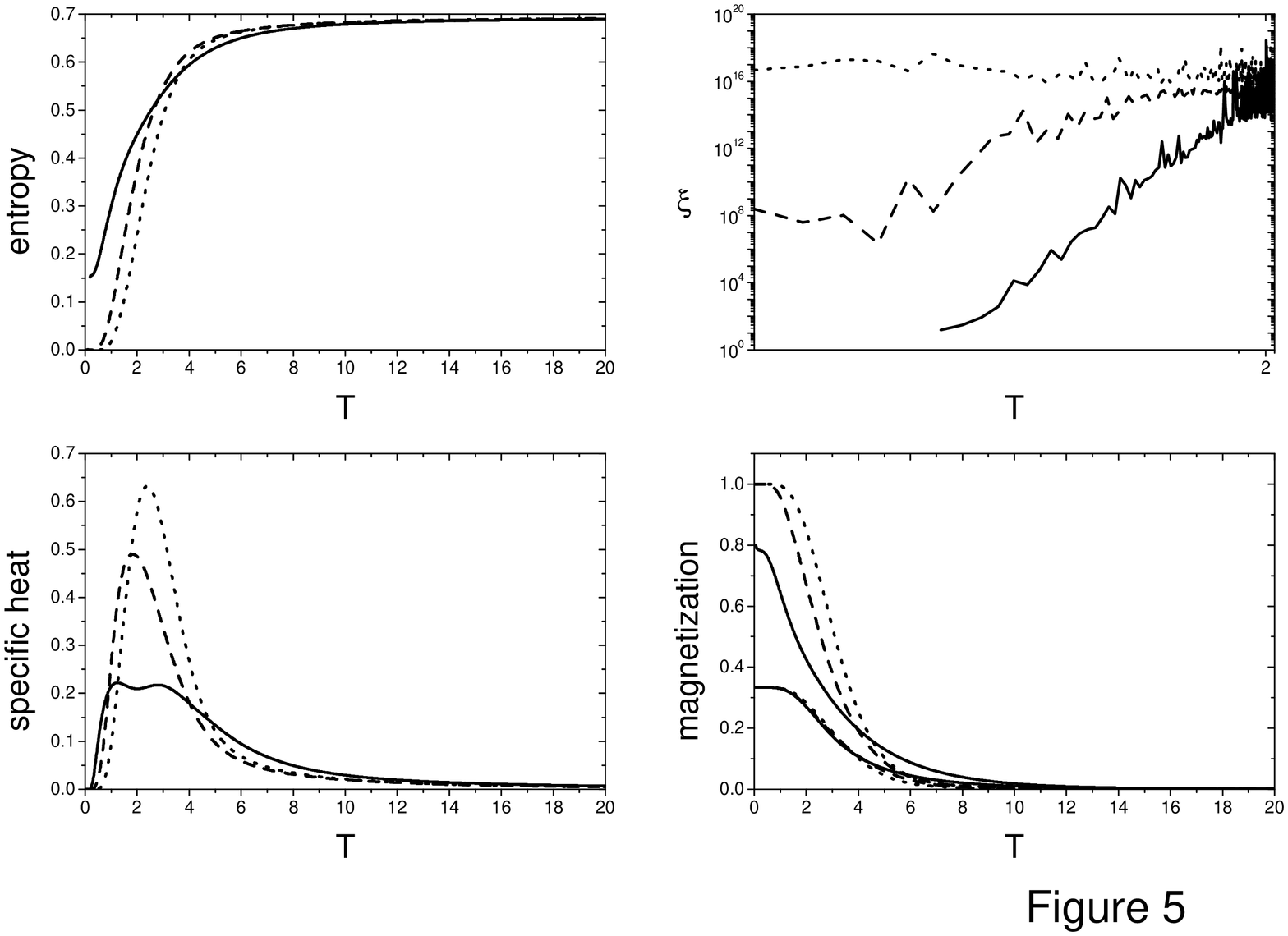}
\end{center}
\caption{Thermodynamic functions for the model with alternating
ferro and anti-ferromagnetic coupling, starting with $J_0=1$. Solid, dash
and dots indicate $\alpha=0,1$ and $\infty$ and $J_0>0$. In (c)
three curves for the magnetization when $J_0=-1$ and same values
of $\alpha$ saturate at $m=1/3$ are also drawn.} \label{fig5}
\end{figure}

In Figure 4 we also draw curves for the thermodynamic functions
when $\alpha=1,$ and $\infty$. In Figure 4c we see that the
magnetization curve always saturates at
$m(T\rightarrow0,h\rightarrow0)=1/3$. This indicates that, in this
limit, the number of spins pointing in opposite directions is not
the same, as is the case for the triangular lattice, but stay  in
proportion 2/3 to 1/3 independent on $\alpha.$ For some range of
values of $\alpha$, $m$ goes through a maximum (e.g. at
$(T,m)=(1.2,0.351)$) when $\alpha=1$, so that a reentrant behavior
at low temperatures is observed.

Let us also discuss how the presence of interactions with
different sign affects the behavior of the system, i.e., when we
take $(-1)^n$ in eq.(\ref{eq8}). As expected, the result depends
whether $J_0>0$ or $<0$. In the first case, competition and
frustration give rise to residual entropy when $\alpha=0$, as
illustrated in Figure 5a. However we note that the value of $s_0$
is smaller $(\sim 0.152)$ than in the case of equal AF
interactions. This happens because not all triangular units are
build by an odd number of AF bonds, as we can easily see by
inspecting the first generations with the help of Figure 1b. Also
for $\alpha=0$ we note the presence of a double Schottky peak in
the specific heat (Figure 5b). For $\alpha\neq0$ there is no
remarkable difference between the curves for $s$ or $c$ with
respect to those obtained for interactions with the same sign. The
results also show that the low temperature magnetization saturates
at the value $m=7/9$ only when $\alpha=0$, otherwise $m=1$.

Finally, when $J_0<0$ and alternating sign are considered, no
frustrated bonds and, consequently, no residual entropy is found.
The curves for the specific heat are also smooth like all other
cases. The magnetization curves saturate again to $m=1/3$ as
$T\rightarrow0$ for all values of $\alpha.$ However, reentrant
behavior similar to that found for some of the AF cases has not
been observed, so that the typical shape is that shown in Figure
5c.

\section{Conclusions}

We have studied a family of Ising models on an Apollonian network
using the transfer matrix technique. On one hand we considered
ferro- and antiferromagnetic couplings and on the other hand we
generalized the interaction as being dependent on the generation
as well in the sign as in the strength quantified by an additional
parameter $\alpha$.

For purely ferromagnetic couplings we always find order in the
thermodynamic limit independent on $\alpha$ which is in agreement
with what has been found on other scale-free
lattices\cite{Stauffer02}. Interestingly the effective critical
temperature at which the correlation length diverges goes to
infinity with the system size with a power-law in the number of
generations with an exponent that depends on $\alpha$. For
antiferromagnetic couplings we find a disordered phase for any
finite temperature but a diverging correlation length at $T=0$.
This later observation is unusual as it does not appear for
instance on the Sierpinski gasket\cite{Andrade93}.

Considering the Apollonian network as a model for the connections
between cities as described in ref.\cite{Andrade04} our  result
can be applied to the formation of opinions where spin up means
one opinion and spin down the other one. The results for the
ferromagnetic case implies that independent on the strength of the
couplings between the cities as long as it is not zero one single
opinion will finally prevail.

If the Appolonian network describes the force lines in a dense
polydisperse packing with each particle having a magnetic moment
as it is the case in tectonic faults\cite{Vilotte} our result for
the ferromagnetic Ising model would imply that if all particles
have a moment of equal strength one always finds a spontaneous
magnetization.

Our calculations can be generalized to random couplings (spin
glass) which in fact is work in preparation. One can also imagine
studying other more complex magnetic models on the Apollonian
networks, like the Potts model, the XY model or the Heisenberg
model and one can also study the magnetic properties of Apollonian
packings of different topology or higher dimension and even the
case of the random Apollonian packing\cite{random}.

\section{Acknowledgement}

H.J. Herrmann thanks the Max Planck Research Award Prize. R.F.S.
Andrade was partially supported by CNPq.

\section{Appendix}

The maps for the free energy and correlation length derived from
eqs. (\ref{eq12}) read:

\begin{equation}
\label{A1}
\begin{array}{l}
  f_{n+1}=\frac{3N_n f_n}{N_{n+1}}-
\frac{T}{N_{n+1}}\{3\ln\alpha_n + \ln\{1+3z_n\beta_n(2 +
\beta_n) + 3z_n^2(1+2\beta_n(1+\beta_n)^2) + \\
z_n^3(1+2\beta_n)(2+\beta_n)^2\} -  6\ln2\},
\end{array}
\end{equation}

\begin{equation}
\label{A2}
\begin{array}{l}
\xi_{n+1}=\xi_n\{1 + \xi_n \\
\{\ln(1+3z_n\beta_n(2+\beta_n)+3z_n^2(1+2\beta_n(1+\beta_n)^2)
+z_n^3(1+2\beta_n)(2+\beta_n)^2)- \\
\ln(z_n^{-1}\beta_n^4 +
z_n(1+4\beta_n+2\beta_n^2+2\beta_n^3) + z_n^2(4+10\beta_n + 11\beta_n^2+2\beta_n^3) + \\
z_n^2(3+10\beta_n+10\beta_n^2+4\beta_n^3))\}\}^{-1}
\end{array}
\end{equation}

where $z_n=(c_n-d_n)/(c_n+d_n)$, $\alpha_n=p_n+q_n$ and
$\beta_n=(p_n-q_n)/(p_n+q_n)$ and $N_n=N(n)$.

\bibliographystyle{prsty}

\end{document}